# Evidence of rf-driven branching of dendritic vortex avalanches in $MgB_2$ microwave resonators

G. Ghigo, F. Laviano, L. Gozzelino, R. Gerbaldo and E. Mezzetti

Department of Physics, Politecnico di Torino, C.so Duca degli Abruzzi 24, 10129 Torino, Italy,

and

Istituto Nazionale di Fisica Nucleare, Sez. Torino, via P. Giuria 1, 10125 Torino, Italy

The influence of dendritic magnetic-flux penetration on the microwave response of superconducting $MgB_2$ films is investigated by a coplanar resonator technique. The peculiar feature consists of jumps in the resonance curve, induced by vortex avalanches freezing flux inside the resonator. Due to a shaking effect, microwave currents maintain the vortex system close to a nonequilibrium state, resulting in dendrite branching. Avalanche-size distributions before and after flux-pinning tailoring by heavy-ion irradiation are fully consistent with molecular dynamics simulations reported in literature.

Avalanche dynamics are found to be major mechanisms in many physical, chemical, biological and social phenomena.[1] They can also be observed in vortex matter when a disordered type-II superconductor is externally driven, for example, by an increasing magnetic field, providing a good test of nonequilibrium dynamics theories. Moreover, since vortex avalanches are usually an undesirable effect for applications, the knowledge of their origin and evolution is of primary importance for device design and development. Concerning the nature of vortex avalanches in superconductors, essentially two different activation mechanisms have been recognized, causing dynamically and thermally driven events. The former is supposed to be the intrinsic mechanism in the formation of the critical state and can be the result of the self organized criticality (SOC) of the system. SOC theories and models predict power-law size distributions for avalanches underlying the dynamics of many systems, with universal exponents.[2] The other kind of flux avalanche, thermally triggered, emerges when a fast increasing of the external field and a small thermal capacity and conductivity of the sample lead to local temperature rise (hot spot), following in a positive feedback the heat dissipation of entering vortices. This causes detachment of vortices from their pinning centers and the resulting instability can destroy the critical state in a large part of the sample. Regarding the theoretical treatment of the issue, vortex avalanche size distributions obtained by molecular dynamics simulations follow power laws with exponents depending on strength and density of pinning sites.[3] Many experimental techniques have been used to investigate these phenomena, e.g. pickup coil detection,[4] SQUID magnetometry,[5] micro Hall probe scanning,[6,7] dc magnetization.[8] Magneto-optical imaging (MOI), which combines high spatial and temporal resolution,[9,10] is probably the most powerful technique since it allows directly imaging the dendritic pattern of flux avalanches.[11,12]

In this Letter we present a detailed study of the influence of dendritic flux penetration on the microwave response of superconducting magnesium diboride ($MgB_2$) resonators, where avalanches are generated by increasing an external dc magnetic field. The peculiar phenomenology consists of characteristic "jumps" in the resonance curves, always leading to a sudden increase of the resonant frequency and of the maximum transmission coefficient, reflecting enhanced rf-current carrying capability. MOI measurements were performed in the absence of the microwave signal in order to give the needed support to the hypothesis that the microwave jumps are induced by vortex avalanches. We have found that once the external field has been stabilized, microwave currents maintain the vortex matter close to a nonequilibrium state, thus causing discontinuous branching of existing avalanches. In fact, in the sand-pile analogy, rf currents would play the role of a "shaking force".[13] Therefore, in this technique, the microwave signal not only allows detecting dendrites, but also contributes to significantly improve the observation statistics, leading to the characterization of the phenomenon in terms of avalanche-size distribution. Since the distribution type may depend on the nature and density of pinning sites, the pinning properties of the same sample have been artificially tailored by high-energy heavy-ion irradiations. Finally, results are compared with the theoretical expectations mentioned above.

Linear coplanar resonators have been obtained by standard photolithography and ion milling from 110 nm-thick $MgB_2$ films on (0001) sapphire substrates.[14] By means of a vector network analyzer we measured the complex transmission coefficient, $S_{21}$, as a function of the driving frequency, $f$, ("resonance curve", henceforth) at different values of temperature and dc external magnetic field applied perpendicular to the film surface.[14]

Magneto-optical images were taken for increasing dc magnetic fields after zero-field cooling down to $T$=5.6K and $T$=8.5K. Irradiations of the resonator took place at the Tandem-XTU accelerator of INFN-LNL, Italy, with 250 MeV Au ions directed perpendicularly to the film surface. We subsequently irradiated the same sample up to partial fluences of $\Phi$=$1.5\cdot10^{11}$ and $\Phi$=$3.0\cdot10^{11}$ $cm^{-2}$. All measurements and irradiations were performed without removing the resonator from the package, in order to avoid any possible change in the mechanical coupling between the strip and the microwave connectors.

Fig.1(a) shows a magneto-optical image of the dendritic flux penetration in the resonator under test, after the application of a dc magnetic field of $H_{appl}$ = 218 Oe at $T$ = 5.6 K. Dendrite formation was observed only at temperatures lower than a threshold, in accordance with literature findings: images taken at T=8.5K did not show any irregular feature. Once the single dendrite has come to a complete expansion, the magnetic flux it contains is practically frozen.[15] This can be clearly seen in figs.1(b)-(c), where images obtained by the difference between two frames taken at consecutive field values are reported. In particular, fig.1(b) was obtained by subtracting the frame taken at $H_{appl}$ = 46 Oe from the frame taken at $H_{appl}$ = 60 Oe. In fig.2(c) the difference of frames taken at $H_{appl}$ = 60 and $H_{appl}$ = 74 Oe is reported for the same region. Dendrites emerging during the first field step (fig.1(b)) are completely frozen during the following field step (fig.1(c)) since they give no significant contribution to the differential image (uniform gray level in the corresponding region, see for example encircled areas). During the second field step new dendrites nucleate from the sample edge, at different starting positions. Moreover, the flux profile inside the dendrite shows a maximum far from the edge, indicating that the screening current at the edges has been recovered after the avalanche expansion.[12] In the following, we show that all the features emerging from MOI measurements have a correspondence in the microwave signal analysis. Moreover, in the latter case, new additional effects emerge from the interaction between flux lines and microwave currents.

The first evidence of an irregular flux penetration at low temperatures in the microwave response of the resonator comes from the fact that the $f_0$ vs. $H$ curve measured at $T$=5K is very noisy if compared to curves measured at higher temperatures (fig.2). A more detailed analysis of single measurements reveals that low-temperature resonance curves are affected by several jumps of various sizes. Such jumps present features characteristic of the dendritic flux penetration: (i) they are irreproducible even for the same resonator in the same conditions (fig.3(a) and (b)); (ii) their number and size depend on the magnetic field variation (fig.3(c)); (iii) they appear only above a threshold field (fig.2) and below a threshold temperature (about 7 K, in this case). Jumps emerge only when the dc field is increased, and very different curves are obtained depending on the measurement starting time (compare curves in fig.3(a), measured just after the dc field increase, with curves in fig.3(b), measured after a delay of 2s).

Once stated the origin of the jumps observed in the microwave response, we must notice that, unlike what happens with other techniques, such jumps always reveal improvements of the resonator performance following the dendritic event. In fact, jumps always cause an increase of the resonant frequency for a forward frequency sweep and an increase of $S_{21}^{max}$ (fig.4). This fact suggests that: (i) after the dendrite expansion towards the sample center the currents at the edge are fully recovered; (ii) the displacement of a certain amount of magnetic flux from the edge to the center is favorable for the resonator since its performance is mostly determined by the conditions at the edge (this is straightforward for the coplanar geometry)[16]; (iii) the flux in the dendrite is frozen and does not contribute significantly to dissipation induced by flux motion. These statements also match the observations emerging from MOI, described above.

As stated in the introduction, the interaction between the microwave signal and relaxing vortices results in dendrite branching. In fact, around the main dendrite (which contains magnetic flux in a sort of liquid state) a contour of "regular" fluxons rapidly relaxes, up to the formation of a stable vortex-density profile that freezes magnetic flux into the dendrite core. If during this relaxation a bundle of vortices gets enough energy from the rf field, it could nucleate a new branch starting from the main dendrite. The correlation between rf-induced dendrite branching and vortex relaxation is confirmed by the dependence of the features observed in the microwave response on the measurement starting time (fig.3).

Further evidence validating this picture comes from the study as a function of the pinning properties, which define the relaxation process, that we gradually modified by high-energy heavy-ion irradiation. Even though no clear evidence of formation of columnar defects has been reported in literature for $MgB_2$, flux pinning improvements by heavy-ions have been proved.[17] Fig.5(a)-(c) show the distributions of the resonant frequency shift, $\delta f_0$, corresponding to dendrite-induced jumps, for the same resonator before irradiation, after irradiation with a fluence of $\Phi$=1.5·10$^{11}$ cm$^{-2}$ and after a subsequent irradiation up to a total fluence of $\Phi$=3·10$^{11}$ cm$^{-2}$. It can be noticed that the total number of events increases with the irradiation fluence and that the distribution shifts towards smaller-size events. The avalanche size, $s$, can be estimated from the total flux variation, proportional to the magnetic field variation, since measurements were performed in quasi-static conditions with respect to the expected dendrite expansion time.[18] Accordingly, $\delta f_0$ is converted to $\delta H_{appl}$ by considering the field dependence of the resonant frequency shown in fig.2. A correction is made to take into account the field focusing effect due to the close proximity of the ground planes and the central strip,[19] and the effective field shift, $\delta H_{eff}$, is obtained, which in turn is proportional to the avalanche size: the corresponding distribution $P(s)$, i.e. the probability to find an avalanche event involving $s$ vortices, is shown in fig.6 for data collected before irradiation and after the higher-dose irradiation. Curves have been fitted to a power law $P(s) \sim s^{-\beta}$, excluding the smaller-event class, that is underestimated since it is close to the sensitivity limit of the technique, and the cut-off due to the upper limit in the avalanche size determined by the finite size of the sample. We have found exponents $\beta=1.00\pm0.07$ for the unirradiated resonator and $\beta=1.44\pm0.06$ for the same resonator after irradiation, in remarkable agreement with the values reported by Olson et al.[3] in their molecular dynamics simulations, where $\beta$ is shown to increase from about 0.9 to 1.4 as the pinning site strength and density are increased. These results suggest a non-SOC scenario, where improved pinning gradually reduces vortex relaxation, making small-size avalanches favorable, in accordance with the assumed model.

In summary, in the present work a comprehensive experimental study on vortex avalanches in $MgB_2$ microwave resonators have been presented. Evidence that peculiar features (jumps) in the resonance curves are determined by dendritic flux penetration has been found. Avalanches freezing magnetic flux inside the resonator produce abrupt improvements of the resonator performance, due to the adopted coplanar geometry (microwave currents flowing mainly along the edges). The microwave signal not only allows detecting avalanches, but also contributes to maintain the system into a nonequilibrium state, generating dendrite branching. The statistic study of this phenomenon led to the determination of the avalanche size distributions in the same sample for different pinning conditions, obtained by improving pinning through heavy-ion irradiation. As a validation of the adopted data analysis and model assumptions, we have found that such distributions follow power laws with exponents that fully agree with values theoretically expected on the basis of molecular dynamics simulations.

This work is partially supported by the Italian National Institute for Nuclear Physics (INFN) under the project DiSCoLi.

**Figure 1 – Magneto-optical image of the dendritic flux penetration in the coplanar resonator (particular) at $T$=5.6K and $H_{appl}$=218Oe, in absence of the microwave signal (a). Brighter colors correspond to higher magnetic flux density regions, while the black zones inside the sample correspond to regions in Meissner state. Figures (b) and (c) show the difference between two frames of the same region (a particular of the central strip) taken at consecutive field values (61 - 46 Oe and 76 – 61 Oe, respectively). Circles indicate that a dendrite grown during the first field step is frozen during the second field step (uniform gray level in the differential image (c)).**

**Figure 2 – Resonance frequency as a function of external dc magnetic field at $T$=5K and at $T$=8.5K, normalized to the value at $H_{appl}$=0. The dotted line marks the field value corresponding to the first detected jump in the microwave response of the resonator at $T$=5K.**

**Figure 3 – Two resonance curves measured in the same conditions, after an external dc field step from $H_{appl}$=0 to $H_{appl}$=13.4 Oe (a). Several resonance curves measured 2s after the same 0-13.35 Oe field step, each of them in the same starting conditions (b). In the inset (c) the same curve shown in (a) is compared to a curve measured after an external dc field step from 0 to 0.45 Oe (dotted) and to a curve measured after a field step from 12.90 to 13.35 Oe (dashed). All measurements started after a zero-field cooling procedure down to $T$=5 K.**

**Figure 4 – Jumps (arrows) detected at $T$=5K in the microwave response of the resonator (solid symbols), following flux avalanches generated by dc field enhancements from 18.25 to 18.70 Oe (a) and from 51.15 to 51.60 Oe (b). Jumps always lead to improvement of the resonator performance since they mark the shift towards different resonance curves (qualitatively represented by dotted lines) with increasing resonant frequency and $S_{21}^{max}$ for a forward frequency sweep.**

**Figure 5 – Resonance frequency-shift histograms for the same resonator before irradiation (a), after irradiation with a fluence of $\Phi=1.5\cdot10^{11}$ cm$^{-2}$ (b), and after a subsequent irradiation up to a total fluence of $\Phi=3\cdot10^{11}$ cm$^{-2}$ (c). Data were collected at $T$=5K and in external magnetic field increasing from 0 to 93.45 Oe, with 0.45 Oe steps.**

**Figure 6 – Number $N(\delta H_{eff})$ of events with the mean effective field shift $\delta H_{eff}$, before irradiation and after the higher-dose irradiation (logarithmic scales). Data were obtained from the histograms of fig.5. The effective field shift is expected to be proportional to the avalanche size, $s$. Therefore the reported power law fits represent the probability $P(s)$ (not normalized) to detect an avalanche of size $s$.**

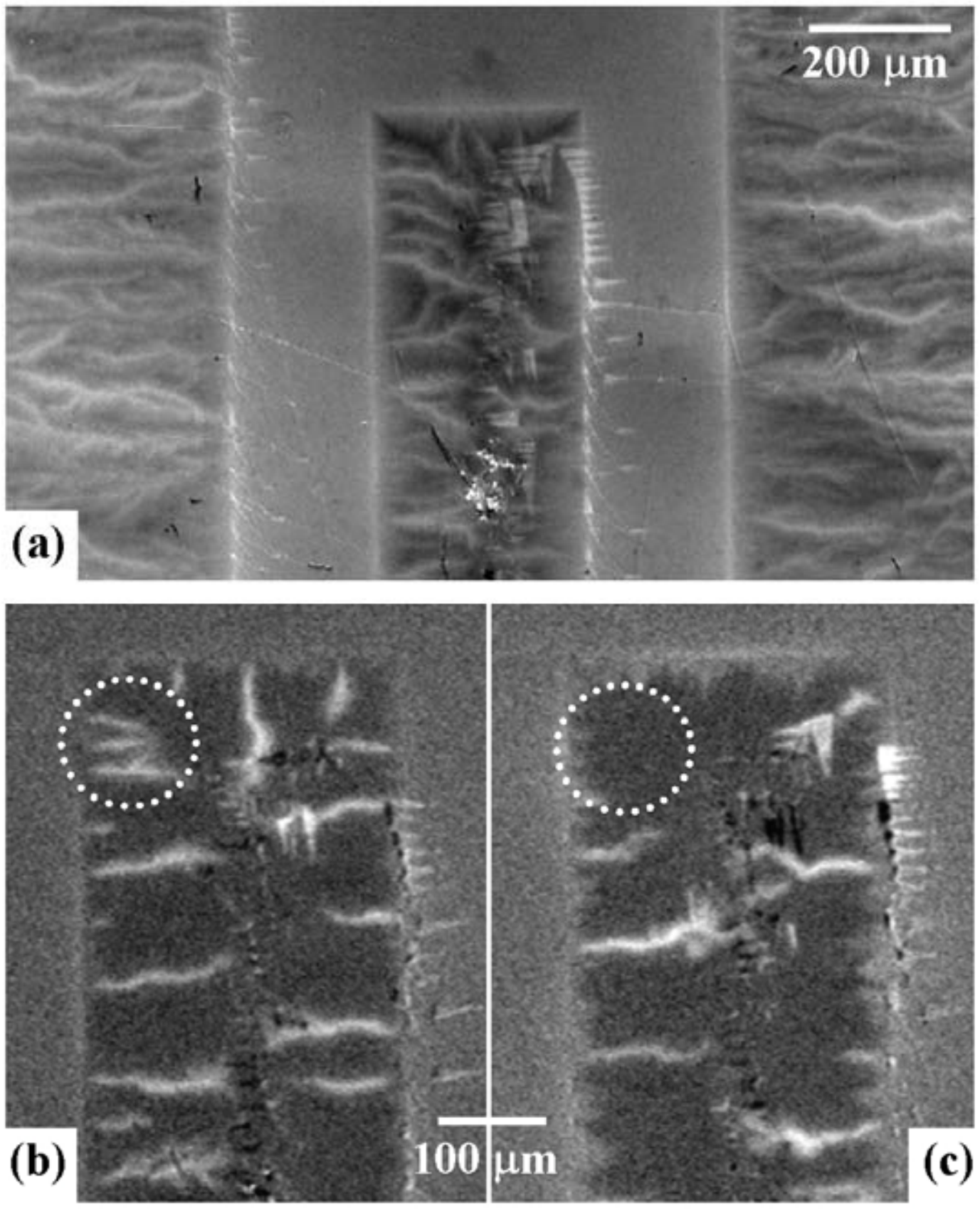

**Figure 1**

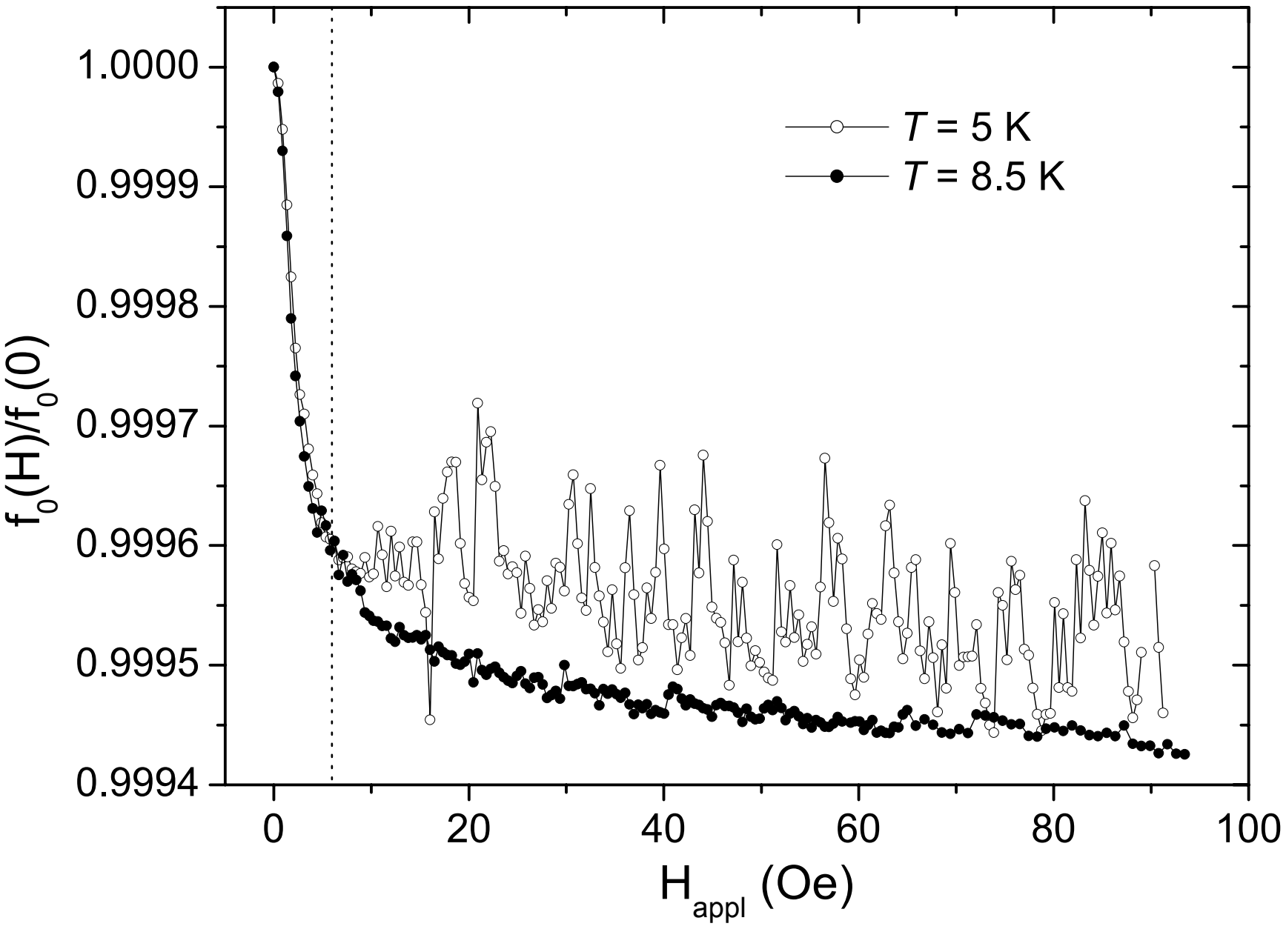

**Figure 2**

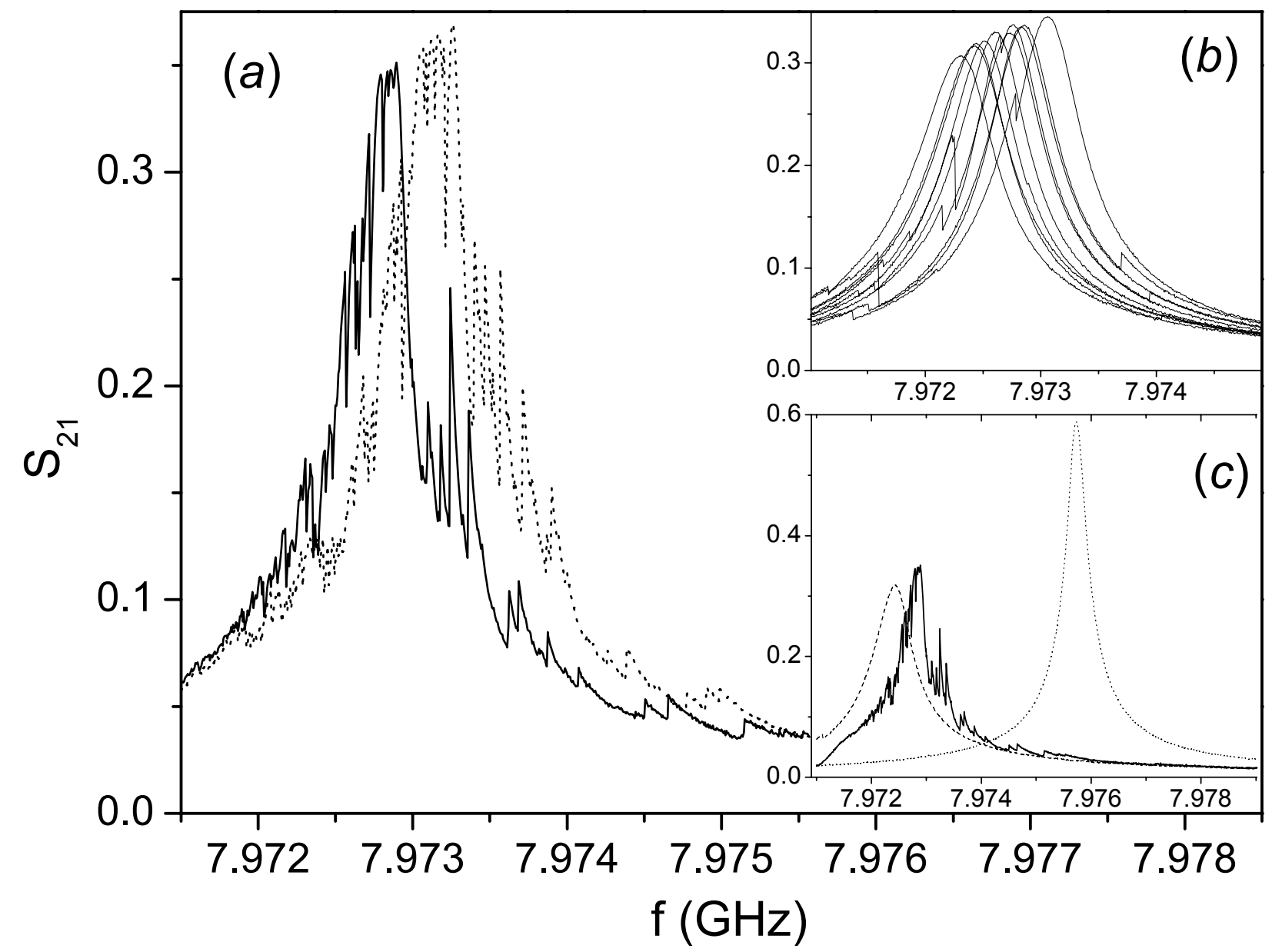

**Figure 3**

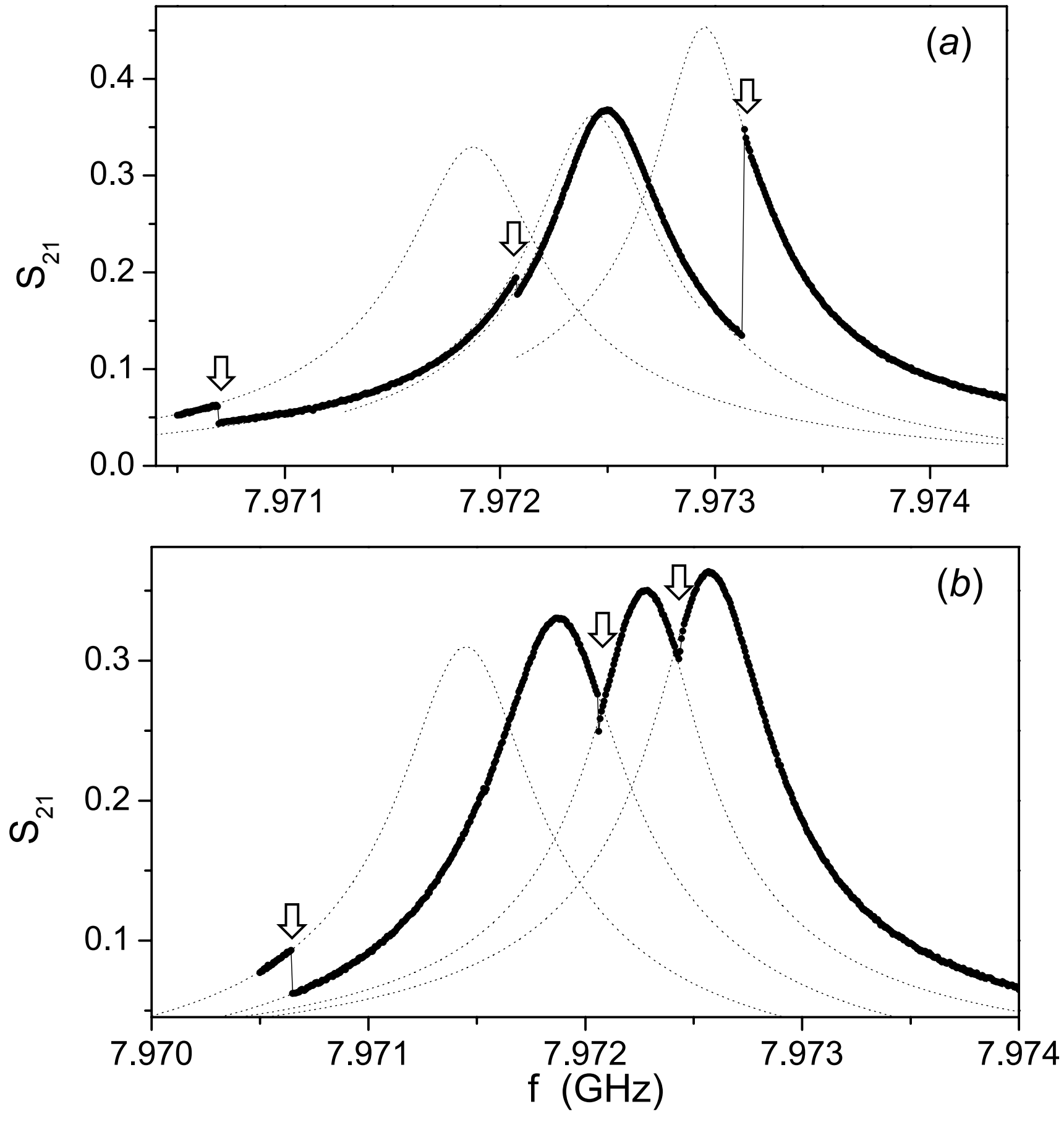

**Figure 4**

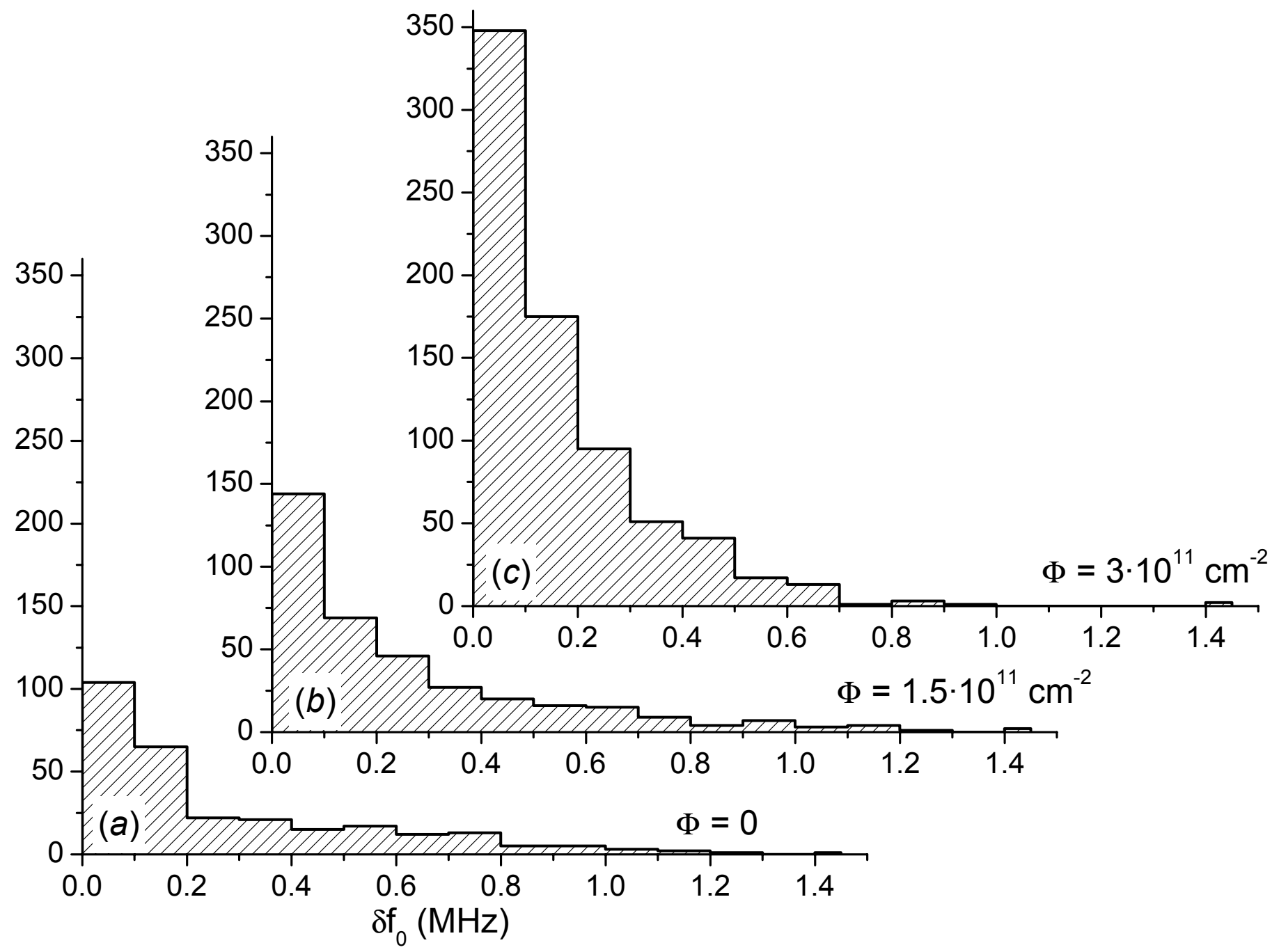

**Figure 5**

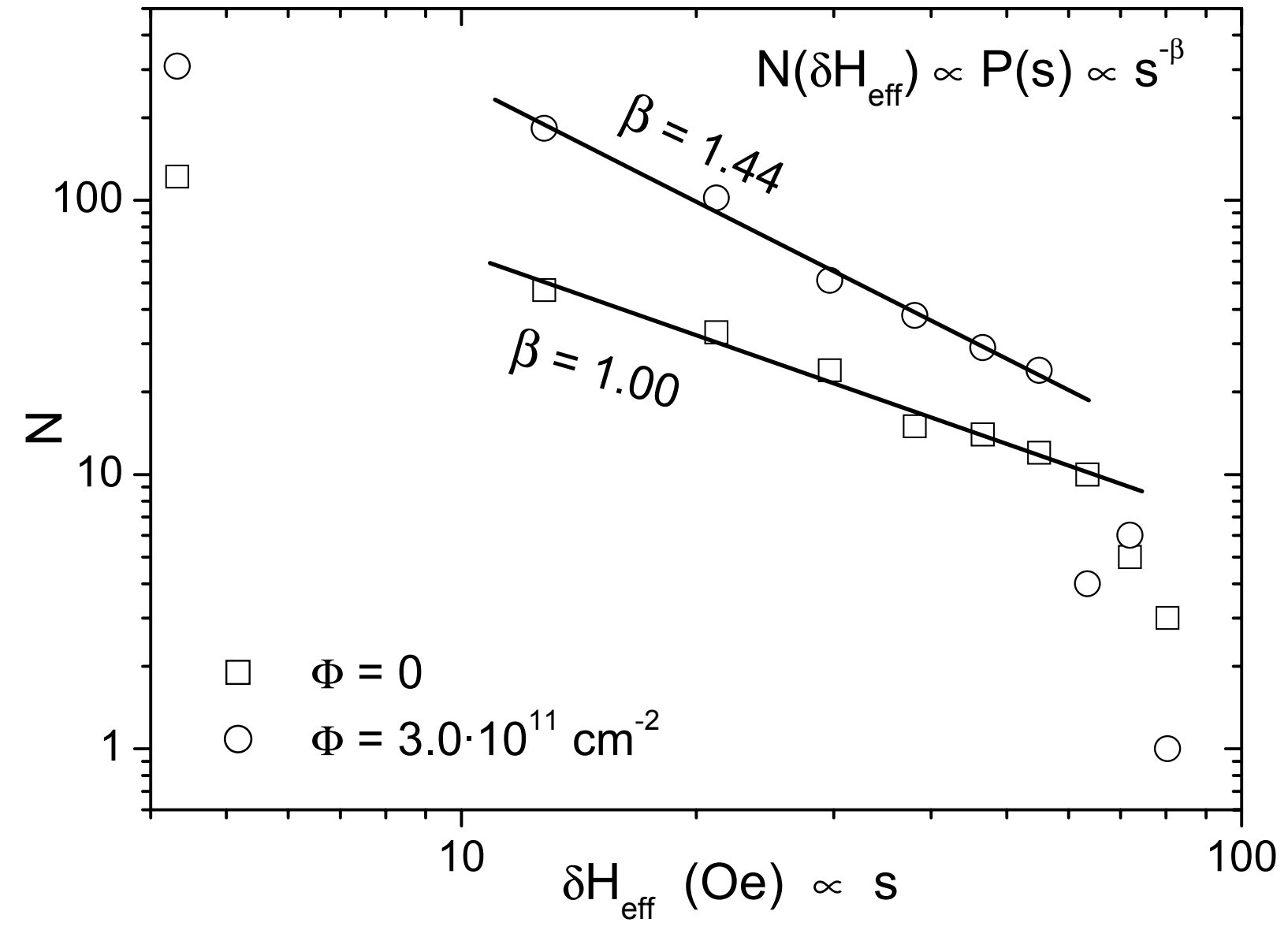

**Figure 6**

[19] Calculations are based on E. H. Brandt and M. Indenbom, Phys. Rev. B **48**, 12893 (1993).